\documentclass[11pt]{article}
\usepackage[utf8]{inputenc}
\usepackage[T1]{fontenc}
\usepackage{graphicx}
\usepackage[bf, tableposition=top]{caption}
\usepackage{longtable}
\usepackage[flushleft]{threeparttable}
\usepackage{ragged2e}
\usepackage[normalem]{ulem}
\usepackage{amsmath}
\usepackage{tikz}
\usepackage{enumitem}
\usepackage{textcomp}
\usepackage{soul}
\usepackage{color}
\usepackage{icomma}
\usepackage{amssymb}
\usepackage{subcaption}
\usepackage{hyperref}
\usepackage{booktabs}
\usetikzlibrary{arrows.meta, bending, positioning, shapes.geometric}

\usepackage{natbib}
\usepackage{lscape}
\usepackage{dcolumn}

\usepackage[onehalfspacing]{setspace} 
\usepackage{parskip}                  

\usepackage[margin=1in]{geometry}
\usepackage{numprint}
\usepackage{adjustbox}
\usepackage{comment}
\usepackage{titling}
\definecolor{darkblue}{rgb}{0.11, 0.27, 0.53} 
\hypersetup{colorlinks = true, urlcolor=darkblue, linkcolor=darkblue, citecolor = darkblue}

\author{Aaron Chatterji\thanks{Aaron Chatterji is the Chief Economist of OpenAI but this
work is not supported by OpenAI and does not necessarily represent its views.}  \\\vspace{.1in} Duke University \and Daniel Rock\thanks{Daniel Rock is grateful to Schmidt Sciences and Kathy Xu for generous financial support.} \\\vspace{.1in}University of Pennsylvania \and Eduard Talamàs\thanks{Eduard Talamàs is grateful to IESE's High Impact Initiative-course 2024/2026 for generous financial support.} \\\vspace{.1in}IESE Business School}

\date{\today}
\title{\vspace{-.3in} The Human-Machine Knowledge Spiral\thanks{We are grateful to Seth Benzell for useful feedback. All errors are our own.}}

\hypersetup{
 pdfauthor={Aaron Chatterji, Daniel Rock, Eduard Talamàs},
 pdftitle={The Human-Machine Knowledge Spiral},
 pdfkeywords={tacit knowledge; artificial intelligence; knowledge creation; organizational learning},
 pdfsubject={Tacit knowledge and artificial intelligence in organizational knowledge creation},
 pdfcreator={LaTeX},
 pdflang={English}}

\begin{document}

\maketitle
\begin{abstract}
Nonaka emphasized that innovation is the result of a continuous back-and-forth between tacit and explicit knowledge. Artificial intelligence introduces a fundamentally new object into this process---\emph{tacit machine knowledge}---but Nonaka's ideas are more relevant than ever. The central role of the knowledge-creating company remains the same: to create the shared context in which different kinds of knowledge can feed off each other, become organizational knowledge, and set off further cycles of innovation.
\end{abstract}

\section{Introduction}

Recent advances in AI are forcing a cognitive version of the Copernican revolution. Just as it once required a profound conceptual leap to accept that the Earth is not the center of the universe, we may now need to accept that human intelligence is not at the center of intelligence \citep{klowdenTao2026}. In organizational terms, this means confronting a possibility that lies outside the assumptions of traditional theories of the firm: human knowledge may no longer occupy a central place in organizational intelligence \citep*{ideTalamas2025,idetalamas2026,chatterjiRockTalamas2025,beraja2026value}. This transformation implicates one of the most influential theories in our field, the knowledge-based view of the firm.  We revisit this theory with this new orientation and argue that Ikujiro Nonaka's insights are more relevant than ever.

Building on \citeauthor{polanyi1966}'s (\citeyear{polanyi1966}) famous observation that ``we know more than we can tell,'' \cite{nonaka1991} conceptualizes innovation as a process of organizational knowledge creation. \emph{Tacit knowledge}---acquired through experience and practice---is articulated into \emph{explicit knowledge} that can be shared across the organization. This explicit knowledge is then internalized through action and use, further expanding tacit knowledge. Through repeated cycles of articulation and internalization, human knowledge is amplified at progressively broader organizational levels---and eventually across organizational boundaries as well. The central role of the knowledge-creating company is to provide the shared context in which this spiral of knowledge can recur.

Artificial intelligence introduces a fundamentally new form of knowledge into organizations \citep*{brynjolfsson2025generative,ideTalamas2025,idetalamas2026,chatterjiRockTalamas2025}. Unlike earlier technologies, AI  learns from data and does not follow explicitly specified rules, so it acquires forms of knowledge that are also difficult to articulate or explain. In other words,  machines now also ``know more than they can tell'' \citep*{brynjolfssonMitchell2017,brynjolfssonMitchellRock2018,autor2024}. Thus, a new type of knowledge has been born: \emph{tacit machine knowledge} \citep*{ideTalamas2025,chatterjiRockTalamas2025}.
Like its human counterpart, tacit machine knowledge is hard to articulate. But unlike its human counterpart, it need not remain local: once embodied in a model, it can be made available at scale through software.\footnote{Machines may not necessarily seek knowledge as ``justified true belief'' as independent agents, but nevertheless can store and act upon information in ways resembling human knowledge processes. \citeauthor{nonaka1994}'s (\citeyear{nonaka1994}) emphasis on the necessity of action, though initially describing human knowledge workers, offers a prescient lens for understanding these AI dynamics.} 

In the age of AI, knowledge no longer moves only between tacit and explicit forms as Nonaka envisioned, but also into and out of machines. Tacit machine knowledge also amplifies the transfer of knowledge across teams and organizations that Nonaka described. But we argue that Nonaka's insights are more relevant than ever. In particular, the central role of the knowledge-creating company continues to be to provide the shared context in which a virtuous spiral of knowledge can recur.

\paragraph{Related literature and contribution.}

This article follows the knowledge-based tradition that asks why firms exist when knowledge is difficult to create, transfer, and use \citep{eppleArgoteDevadas1991,kogutZander1992,zanderKogut1995, kogutZander1996,grant1996,argote1999}. \citeauthor{nonaka1991}'s (\citeyear{nonaka1991}) distinctive contribution is to show that knowledge creation requires making different kinds of knowledge available to one another in a shared context \citep{nonaka1994,nonakaTakeuchi1995,nonakaKonno1998, nonakaVonKrogh2009}.

A growing literature has begun to revisit this literature in light of AI and generative AI. \cite{sanzogniGuzmanBusch2017} connects artificial intelligence to the knowledge-management debate by asking whether AI changes how we should understand the tacit dimension of knowledge. More recent work treats ChatGPT and related systems as tools that can support Nonaka's original spiral of knowledge  \citep{sumbalAmber2025}. Other work goes further and asks how generative AI can be treated as a participant in knowledge creation  \citep{matsumotoNishikawaMorimoto2025,bohmDurst2026,kirchnerScarso2026, uchihira2026}. A related line of work asks whether AI-mediated interaction can itself become a new form of shared context \citep{heAntonczakBurgerHelmchen2026}. This literature is important because it recognizes that AI can participate in the processes through which organizations create knowledge.

The novelty of this paper is to revisit Nonaka's ideas in the context of the rise of a fundamentally new form of knowledge---tacit machine knowledge---which does not sit comfortably in the traditional tacit--explicit divide that \cite{polanyi1966} emphasizes and \cite{nonaka1991} builds on. We argue that Nonaka's insights and prescriptions apply \emph{mutatis mutandis} once the relevant spiral includes tacit machine knowledge. The key property of this new type of tacit knowledge is portability: it can be deployed at scale while remaining opaque to humans. AI therefore dramatically amplifies the transfer across teams and organizational boundaries that Nonaka emphasizes, as well as the importance of shared context for the knowledge-creating company.

\section{The Human--Machine Knowledge Spiral} 

\cite{nonaka1991} emphasizes that making personal knowledge available to others is the central activity of the knowledge-creating company. As the following example illustrates, in the age of AI, a new but analogous activity becomes central for innovation: to make human and machine knowledge available to one another. 

In 2015, Matías Muchnick, Karim Pichara, and Pablo Zamora founded NotCo with an ambitious goal: to reproduce familiar animal-based foods using plants. They built an artificial-intelligence system called Giuseppe, after the Renaissance painter Giuseppe Arcimboldo, who composed human faces out of fruits and vegetables.\footnote{See ``Chef Giuseppe Heralds a New Culinary Era,'' \href{https://www.unesco.org/en/articles/chef-giuseppe-heralds-new-culinary-era-0}{UNESCO Courier}, June 25, 2018 (updated May 12, 2023) and ``The Evolution of Giuseppe, an AI System That Generates Fake Meat and Dairy Recipes,'' \href{https://www.techbrew.com/stories/2021/08/18/evolution-giuseppe-ai-system-generates-fake-meat-dairy-recipes}{Tech Brew}, August 18, 2021.  }

Giuseppe searched across ingredient data, chemical analyses, scientific publications, previous formulations, and sensory results on a scale no food scientist could match. Given a target such as milk, it proposed plant combinations that might reproduce its flavor, aroma, texture, appearance, nutrition, and behavior in cooking. Because Giuseppe could explore a vastly larger space than any human team, it was tempting to think that NotCo's creative problem had been solved.

Experience proved otherwise. In one early attempt to create NotMilk, Giuseppe proposed a formulation containing dill. The milk came out green. Giuseppe had detected properties in dill that appeared useful in the data, but it missed something obvious to anyone standing in NotCo's kitchen: whatever else milk may be, people do not expect it to be green. The failure made it obvious to everyone involved that---despite having access to large amounts of data---Giuseppe had a serious lack of context.\footnote{See ``Let's take a look at how AI is changing food production'' \href{https://www.foodprocessing.com.au/content/food-design-research/article/let-s-take-a-look-at-how-ai-is-changing-food-production-1385714050}{Food Processing}, December 9, 2022.}

The green milk vividly illustrates the different nature of machine knowledge. Giuseppe excels at detecting potentially useful relationships among ingredients and molecular compounds. But it cannot taste the product, smell it, or understand everything people expect from a glass of milk. In contrast, chefs can recognize if a prototype is too grassy, too thin, too sweet, or simply wrong. They can appreciate color, aroma, mouthfeel, aftertaste, stability, and performance under heat. Much of this knowledge has been acquired through experience and is difficult to reduce to rules. 

NotCo's response to the green milk fiasco was to orchestrate a spiral of knowledge between Giuseppe and its human teams. Giuseppe proposed formulas. Research chefs made them. Chefs, food scientists, and sensory panels tasted, smelled, measured, and modified the results. Their judgments about color, flavor, texture, mouthfeel, and appearance were recorded and returned to Giuseppe, which then generated new proposals in light of the feedback.\footnote{See NotCo's patent ``Systems and methods to mimic target food items using artificial intelligence,'' \href{https://patents.google.com/patent/US11164478B2/en}{U.S. Patent No. 11,164,478 B2} and ``Formula and recipe generation with feedback loop,'' \href{https://patents.google.com/patent/US11205101B1/en}{U.S. Patent No. 11,205,101 B1}. See also ``The Evolution of Giuseppe, an AI System That Generates Fake Meat and Dairy Recipes,'' \href{https://www.techbrew.com/stories/2021/08/18/evolution-giuseppe-ai-system-generates-fake-meat-dairy-recipes}{Tech Brew}, August 18, 2021.}  By 2021, more than twenty chefs and food scientists were working with Giuseppe---testing more than one hundred recipes a month.\footnote{See ``The evolution of Giuseppe, an AI system that generates fake meat and dairy recipes'' \href{https://www.techbrew.com/stories/2021/08/18/evolution-giuseppe-ai-system-generates-fake-meat-dairy-recipes}{Tech Brew}, August 18, 2021.} 

Many proposals failed. Others opened paths that the human teams were unlikely to have explored. In searching for a plant formulation with the aromatic qualities of cow's milk, Giuseppe helped direct the team toward an improbable but winning combination: pineapple and cabbage.\footnote{See ``Inside NotCo’s plant-based empire'' \href{https://www.foodbusinessnews.net/articles/18463-inside-notcos-plant-based-empire}{Food Business News}, April 26, 2021.}  NotMilk---with its pineapple and cabbage---reached Whole Foods stores nationwide in the United States in 2020.\footnote{See ``Bezos backed NotCo launches NotMilk™ in the United States at Whole Foods Market Stores'' \href{https://www.businesswire.com/news/home/20201102005108/en/Bezos-backed-NotCo-launches-NotMilk-in-the-United-States-at-Whole-Foods-Market-Stores}{Business Wire}, November 2, 2020.}  A 2021 financing round valued NotCo at \$1.5 billion.\footnote{See ``Food Tech Pioneer NotCo Announces \$235M Series D Round at \$1.5 Billion Valuation'' \href{https://www.businesswire.com/news/home/20210726005214/en/Food-Tech-Pioneer-NotCo-Announces-235M-Series-D-Round-at-1.5-Billion-Valuation}{Business Wire, Series D}, July 26, 2021.}

Nonaka emphasized that products embody organizational knowledge. They are the concrete forms in which organizational knowledge is crystallized, and they are also triggers that set off the next cycle of knowledge creation \citep{nonaka1991,nonaka1994}. Matsushita's bread machine carried the baker's ``twist dough'' insight beyond the team that had created it. The machine made that insight available outside the original process in a durable form. In this sense, the spiral of knowledge in one team can help set further spirals in motion in other teams.\footnote{For a critical analysis of the stronger claim that the baker's tacit knowledge was made explicit and incorporated into the bread-making machine, see \cite{ribeiroCollins2007}. Our point does not require that stronger knowledge-capture thesis. The weaker Nonaka point is enough: product-development work can crystallize knowledge in a durable artifact and trigger further knowledge creation.}
 
Similarly, NotMilk embodies the formula created through NotCo's human-machine knowledge spiral. Its ingredient list makes the pineapple-and-cabbage combination visible. Other consumers and companies can taste the result, inspect it, and learn from it.

But NotCo can now make a second object travel: Giuseppe itself. 

Giuseppe embodies tacit machine knowledge accumulated through years of formulas, sensory evaluations, failures, and corrections. This knowledge cannot be articulated as a set of explicit rules, but can be transferred at scale through software. 

Indeed, this possibility has become part of NotCo's business. In 2022, the company raised \$70 million to develop a business-to-business unit through which other consumer-goods companies could use Giuseppe in their own product development. NotCo continues to sell food, while NotCo AI offers Giuseppe's tacit knowledge as an innovation platform for other firms.\footnote{See ``Global Food Tech Unicorn NotCo Closes \$70M in Funding to Fuel New B2B Platform'' \href{https://www.businesswire.com/news/home/20221212005094/en/Global-Food-Tech-Unicorn-NotCo-Closes-70M-in-Funding-to-Fuel-New-B2B-Platform}{Business Wire}, December 12, 2022. See also \href{https://www.notco.ai/}{NotCo AI}.} This is Nonaka's movement of knowledge across organizations---at a new scale. 


For example, in 2025, NotCo AI announced a partnership with the chocolate maker Barry Callebaut. The announced partnership is intended to combine NotCo AI's learned capabilities with Barry Callebaut's century of chocolate expertise, ingredient knowledge, manufacturing infrastructure, supply chain, and distribution infrastructure. In other words, Giuseppe is being moved into a new domain: from the plant-based milk problem that made NotCo famous to the development of chocolate recipes.\footnote{See ``Barry Callebaut Partners with NotCo AI to Unlock Next-Level Chocolate Innovation,'' \href{https://www.barry-callebaut.com/en/about-us/media/news-stories/barry-callebaut-partners-notco-ai-unlock-next-level-chocolate}{Barry Callebaut}, November 18, 2025.  See also ``Barry Callebaut to Use NotCo AI to Develop Chocolate Recipes,'' \href{https://www.reuters.com/business/barry-callebaut-use-notco-ai-develop-chocolate-recipes-2025-11-18/}{Reuters}, November 18, 2025, and \href{https://www.notco.ai/}{NotCo AI}.}

NotCo's case suggests that making different types of knowledge available to one another continues to be the central activity of the knowledge-creating company. The distinctive organizational problem is that the relevant spiral now includes tacit machine knowledge. This adds five movements to Nonaka's four modes of knowledge conversion (socialization, externalization, combination, and internalization). The first four connect human and machine knowledge. The fifth connects machine to machine---and highlights why tacit machine knowledge can scale beyond the local setting in which it was created.

\paragraph{From explicit knowledge to tacit machine knowledge.}

An organization can incorporate documents, formulas, scientific relationships, technical specifications, and experimental records into a learning system. Machines can use this material to form internal relationships that shape what they can recognize and propose. We call this process \emph{encoding}. At NotCo, ingredient specifications, chemical measurements, prior recipes, and production constraints become part of Giuseppe's learned capabilities. Yet encoding alone is limited: a machine may absorb everything an organization has written down and still miss what its members take for granted---like the fact that milk is not green. 

\paragraph{From tacit human knowledge to tacit machine knowledge.}

Human tacit knowledge can be partially transferred to machines through choices, demonstrations, corrections, ratings, and modifications. We call this process \emph{embedding}. This process is akin to \emph{socialization} of machines. By repeatedly distinguishing better from worse formulations, NotCo's chefs make part of their tacit knowledge available to Giuseppe.

\paragraph{From tacit machine knowledge to explicit human knowledge.}

Machines can express implications of what they have learned in the form of predictions, designs, hypotheses, or recipes. We call this process \emph{extraction}, analogous to Nonaka's \emph{externalization} when done from human tacit knowledge to human explicit knowledge. A formula containing pineapple and cabbage turned an inaccessible machine insight into an explicit proposal that people could make, examine, and test. More generally, Giuseppe makes part of its tacit knowledge available to NotCo's human teams through its wild recipe proposals---regardless of whether they are successful.

\paragraph{From tacit machine knowledge to tacit human knowledge.}

People can also learn from machines through observation, experimentation, and practice. We call this process \emph{assimilation}. When a chef repeatedly makes and refines combinations she would never have considered on her own, Giuseppe's insights begin to alter what she notices, expects, and imagines. Assimilation is the mirror image of embedding: tacit human knowledge shapes machine knowledge, and tacit machine knowledge reshapes human knowledge.

\paragraph{From tacit machine knowledge to tacit machine knowledge}

Machine tacit knowledge is expensive to create, but the marginal cost of transferring it to other machines is relatively low. Once knowledge is embodied in model weights, architectures, embeddings, memories, or other machine-readable representations, it can be copied, distilled, fine-tuned, deployed through an API, or incorporated into another system. This differs substantially from tacit human knowledge. Human tacit knowledge usually travels slowly through interpersonal interaction. Tacit machine knowledge can remain tacit while moving at software speed. A model trained in one organization can therefore become the starting point for another model, product, workflow, or organization. This gives the human-machine knowledge spiral a new scaling mechanism: knowledge created in one local setting can be replicated and recombined across many other settings without ever being fully articulated. We call this \emph{machine transfer}. When NotCo creates copies or new versions of Giuseppe, there is no need to bring Giuseppe's latent knowledge to human collaborators first.

NotCo illustrates how---when these five knowledge movements feed off each other---something powerful happens:

\begin{enumerate}
    \item First, the company encodes scientific knowledge, ingredient data, recipes, and experimental records into Giuseppe.
    \item Next, chefs and food scientists embed aspects of their tacit judgment through repeated evaluations and corrections.
    \item Giuseppe then extracts its tacit insights in the form of new formulas that human teams can test.
    \item By making and refining Giuseppe's wild formulas, human teams assimilate new possibilities into their own tacit knowledge. 
    \item Finally, through machine transfer, Giuseppe's learned capabilities can contribute to new systems, partner workflows, and product-development contexts beyond the original NotCo team.
   
\end{enumerate}

This starts the spiral all over again, but at a higher level. The human teams' enlarged tacit knowledge produces new experiments and evaluations. These create new explicit records and new traces of human experience. Both return to the machine, which generates new possibilities for extraction and assimilation. Through machine transfer, the resulting tacit machine knowledge can also seed other spirals before it has been fully articulated.

\citeauthor{nonaka1991}'s (\citeyear{nonaka1991}) original spiral of knowledge between tacit and explicit knowledge remains relevant. Chefs still learn from one another through practice, articulate discoveries, combine explicit knowledge, and internalize it through action. But these conversions now interact with the second spiral connecting people and machines---and this is the human-machine knowledge spiral that the knowledge-creating company must deliberately orchestrate and nurture in the age of AI.

\section{From Chaos to Concept---Again}

\cite{nonaka1991} emphasizes the value of \emph{redundancy}: the conscious overlap of information, activity, and responsibility. Redundancy creates the common cognitive ground on which people situated in different contexts can sense what others are struggling to articulate.  It places different participants around the same problem and allows their different perspectives to interact.

At NotCo, Giuseppe, the chefs, the food scientists, and the sensory panels repeatedly confronted the same prototype. The machine examined relationships among ingredients and molecular properties. The chefs experienced its color, aroma, taste, texture, mouthfeel, and culinary behavior. Food scientists measured stability, nutrition, and production performance. Their work overlapped at the prototype on the kitchen table.

The prototype was a shared context---what \cite{nonakaKonno1998} call \emph{ba}---in which knowledge could acquire meaning through interaction. From the standpoint of information processing, having multiple expert chefs carefully evaluating Giuseppe's formulas may look like duplication. From the standpoint of knowledge creation, this redundancy creates the common cognitive ground whose value Nonaka emphasized.\footnote{See ``NotCo Built a Unicorn Using AI To Accelerate Food Innovation. CEO Matias Muchnick Tells The Spoon How They Did It,'' \href{https://thespoon.tech/notco-built-a-unicorn-using-ai-to-accelerate-food-innovation-ceo-matias-muchnick-tells-the-spoon-how-they-did-it/}{The Spoon}, March 3, 2022.}

Discrepancies are especially valuable. Nonaka argued that ambiguity, conflict, and breakdown can force members of an organization to reexamine what they take for granted. The green milk did exactly that. The fiasco exposed the distance between molecular resemblance and the lived meaning of milk, and made part of the chefs' tacit knowledge visible: milk has a color, and that color belongs to what people understand milk to be.\footnote{See ``Let's Take a Look at How AI Is Changing Food Production,'' \href{https://www.foodprocessing.com.au/content/food-design-research/article/let-s-take-a-look-at-how-ai-is-changing-food-production-1385714050}{Food Processing}, December 9, 2022.}

Giuseppe's successful proposals also echo Nonaka's movement from metaphor to model. ``Pineapple and cabbage as milk'' brought distant domains together before anyone possessed a complete explanation of their connection. The chefs and scientists had to examine precisely how these ingredients were and were not like the product they hoped to reproduce. A wild proposal turned into useful shared knowledge through a process that sounds very similar to the one Nonaka described.\footnote{See  ``Inside NotCo's Plant-Based Empire,'' \href{https://www.foodbusinessnews.net/articles/18463-inside-notcos-plant-based-empire}{Food Business News}, April 26, 2021.}

Knowledge creation also requires direction. Nonaka insisted that the knowledge-creating company is as much about ideals as it is about ideas. Giuseppe could search what was possible. NotCo's ideal---familiar food made from plants without surrendering the qualities people valued---gave that search a purpose. The ideal does not specify dill, pineapple, or cabbage. Instead, it gives the team a way to decide which wild proposals deserve another turn of the spiral.\footnote{See ``Chef Giuseppe Heralds a New Culinary Era,'' \href{https://www.unesco.org/en/articles/chef-giuseppe-heralds-new-culinary-era-0}{UNESCO Courier}, June 25, 2018, updated May 12, 2023. See also  \href{https://cdn2.notco.com/uploads/NotCo_FS_Brochure.pdf}{NotCo Foodservice Brochure}.}

Nonaka's account of organizational roles reads with similar force today. Frontline employees possess detailed knowledge of ``what is.'' Senior leaders give voice to ``what ought to be.'' Middle managers operate between the two, translating broad ideals into concepts that can encounter the reality of products, technologies, and markets. They are, in \citeauthor{nonaka1991}'s (\citeyear{nonaka1991}) words, the knowledge engineers of the company.

AI adds another source of knowledge to this structure. Domain experts understand the reality of the product and the customer, AI specialists know AI's capabilities, and managers have a sense of where these can be useful. The machine contributes patterns and proposals that neither group may have anticipated. The work of connecting these perspectives to one another and to the organization's purpose is analogous to the work Nonaka emphasized was central to the knowledge-creating company.

Nonaka's insights also suggest that knowledge creation cannot reside inside an isolated AI unit. Nonaka gives no department or group of experts exclusive responsibility for creating knowledge. Senior managers, middle managers, and frontline employees all participate. Their contributions matter because of what they add to the knowledge-creating system as a whole. In the age of AI, the same principle includes model builders, domain experts, operating employees, and the machines themselves. 

The importance of shared context becomes clearest when tacit machine knowledge crosses organizational boundaries via machine transfer. The partnership announced in 2025 between NotCo AI and Barry Callebaut can be read in these terms. From Nonaka's perspective, the partnership has the elements of a new \emph{ba}: a setting in which NotCo's portable machine knowledge and Barry Callebaut's situated knowledge can acquire meaning through interaction. Giuseppe can supply possibilities, while chocolate experts make, taste, interpret, and revise them. If the process follows the feedback logic described in NotCo's own patents and interviews, their responses can create new knowledge that returns to Giuseppe. In this sense, the spiral created at NotCo can feed Barry Callebaut's spiral, which may continue the innovation process in its own context.

\section{Conclusion}

While everything about Giuseppe would have been unimaginable when Nonaka wrote about Matsushita's bread machine 35 years ago, we believe the organizational process shaping it would have been very familiar to him. An insight appears in one place, encounters a different form of knowledge, acquires meaning in a shared context, and returns to begin another cycle of the knowledge spiral.

The AI revolution itself offers another illustration of how relevant Nonaka's insights are today. Before ChatGPT could converse with the world, human trainers staged conversations in which they played both the user and the assistant. They were given model-written suggestions to help compose their responses. They then ranked alternative replies generated by the model, and those judgments shaped subsequent training \citep{ouyangEtAl2022}.\footnote{See ``Introducing ChatGPT'' \href{https://openai.com/index/chatgpt/}{OpenAI}, November 30, 2022.}

The same movements seen at NotCo appear here at a different scale. Vast bodies of explicit text are encoded in the model. Human demonstrations and rankings embed judgments about what constitutes a useful response. Model outputs make learned capabilities available for inspection and evaluation. By observing unexpected successes and failures, the people building and using the system acquire new knowledge about what the model can do and where it remains limited. 

This is the human--machine knowledge spiral operating inside an AI laboratory. The main product of this spiral is tacit machine knowledge that can be made available far beyond the laboratory. Through software products and APIs, tacit machine knowledge formed inside one laboratory is now becoming central to people and organizations around the world \citep{chatterji2025people,johnston2026shift}.  The spiral operating inside a small number of AI laboratories is therefore feeding the knowledge spirals of countless other companies---at scale.

Here again, Nonaka explains why access to tacit machine knowledge alone---no matter how impressive---does not create organizational knowledge. AI models arrive to an organization without its context or ideals. The imported knowledge becomes organizational knowledge only after it enters the human-machine knowledge spiral.   The rise of AI will transform firms, but the knowledge-creating company still has to provide the shared context in which a virtuous spiral of knowledge can recur.

{\linespread{1.3}
\bibliographystyle{ecta}
\bibliography{nonaka}}

@article{sanzogniGuzmanBusch2017,
  author  = {Sanzogni, Louis and Guzman, Gustavo and Busch, Peter},
  title   = {Artificial Intelligence and Knowledge Management: Questioning the Tacit Dimension},
  journal = {Prometheus},
  year    = {2017},
  volume  = {35},
  number  = {1},
  pages   = {37--56},
  doi     = {10.1080/08109028.2017.1364547}
}

@article{ribeiroCollins2007,
  author  = {Ribeiro, Rodrigo and Collins, Harry},
  title   = {The Bread-Making Machine: Tacit Knowledge and Two Types of Action},
  journal = {Organization Studies},
  year    = {2007},
  volume  = {28},
  number  = {9},
  pages   = {1417--1433},
  doi     = {10.1177/0170840607082228}
}

@article{sumbalAmber2025,
author = {Sumbal, Muhammad Saleem and Amber, Quratulain},
title = {{ChatGPT}: A Game Changer for Knowledge Management in Organizations},
journal = {Kybernetes},
year = {2025},
volume = {54},
number = {6},
pages = {3217--3237},
doi = {10.1108/K-06-2023-1126},
url = {https://doi.org/10.1108/K-06-2023-1126}
}

@article{bohmDurst2026,
author = {Böhm, Karsten and Durst, Susanne},
title = {Knowledge Management in the Age of Generative Artificial Intelligence: From {SECI} to {GRAI}},
journal = {VINE Journal of Information and Knowledge Management Systems},
year = {2026},
volume = {56},
number = {1},
pages = {106--121},
doi = {10.1108/VJIKMS-10-2024-0357},
url = {https://doi.org/10.1108/VJIKMS-10-2024-0357}
}

@incollection{kirchnerScarso2026,
author = {Kirchner, Kathrin and Scarso, Enrico},
title = {Revisiting the {SECI} Model: The Impact of Generative {AI} on Organizational Knowledge Creation},
booktitle = {Managing Human and Artificial Knowledge},
editor = {Bolisani, Ettore and Nakash, Maayan and Bratianu, Constantin and Bejinaru, Ruxandra},
series = {Knowledge Management and Organizational Learning},
volume = {17},
publisher = {Springer},
address = {Cham},
year = {2026},
pages = {105--125},
doi = {10.1007/978-3-032-14721-9_6},
url = {https://doi.org/10.1007/978-3-032-14721-9_6}
}

@misc{uchihira2026,
author = {Uchihira, Naoshi},
title = {Tacit Knowledge Management with Generative {AI}: Proposal of the {GenAI SECI} Model},
year = {2026},
note= {Working Paper},
eprint = {2603.21866},
archivePrefix = {arXiv},
primaryClass = {cs.AI},
doi = {10.48550/arXiv.2603.21866},
url = {https://arxiv.org/abs/2603.21866}
}

@incollection{matsumotoNishikawaMorimoto2025,
author = {Matsumoto, Takashi and Nishikawa, Ryu and Morimoto, Chikako},
title = {Human-{AI}-Collaboration {SECI} Model: The Knowledge Management Model of the Experts' Tacit Knowledges with Augmented {LLM}-Based {AI}},
booktitle = {Agents and Multi-agent Systems: Technologies and Applications 2024},
editor = {Jezic, Gordan and Chen-Burger, Y.-H. and Ku{\v{s}}ek, Mario and {\v{S}}perka, Roman and Howlett, Robert J. and Jain, Lakhmi C.},
series = {Smart Innovation, Systems and Technologies},
volume = {406},
publisher = {Springer},
address = {Singapore},
year = {2025},
pages = {135--145},
doi = {10.1007/978-981-97-6469-3_12},
url = {https://doi.org/10.1007/978-981-97-6469-3_12}
}

@article{heAntonczakBurgerHelmchen2026,
author = {He, Xiaomei and Antonczak, Laurent and Burger-Helmchen, Thierry},
title = {Revisiting {Ba}: How Generative {AI} Transforms Knowledge Creation},
journal = {Journal of Knowledge Management},
year = {2026},
volume = {30},
number = {2},
pages = {846--869},
doi = {10.1108/JKM-07-2025-1065},
url = {https://doi.org/10.1108/JKM-07-2025-1065}
}

@article{eppleArgoteDevadas1991,
author = {Epple, Dennis and Argote, Linda and Devadas, Rukmini},
title = {Organizational Learning Curves: A Method for Investigating Intra-Plant Transfer of Knowledge Acquired Through Learning by Doing},
journal = {Organization Science},
year = {1991},
volume = {2},
number = {1},
pages = {58--70},
doi = {10.1287/orsc.2.1.58},
url = {https://doi.org/10.1287/orsc.2.1.58}
}

@article{kogutZander1992,
author = {Kogut, Bruce and Zander, Udo},
title = {Knowledge of the Firm, Combinative Capabilities, and the Replication of Technology},
journal = {Organization Science},
year = {1992},
volume = {3},
number = {3},
pages = {383--397},
doi = {10.1287/orsc.3.3.383},
url = {https://doi.org/10.1287/orsc.3.3.383}
}

@article{zanderKogut1995,
author = {Zander, Udo and Kogut, Bruce},
title = {Knowledge and the Speed of the Transfer and Imitation of Organizational Capabilities: An Empirical Test},
journal = {Organization Science},
year = {1995},
volume = {6},
number = {1},
pages = {76--92},
doi = {10.1287/orsc.6.1.76},
url = {https://doi.org/10.1287/orsc.6.1.76}
}

@article{kogutZander1996,
author = {Kogut, Bruce and Zander, Udo},
title = {What Firms Do? Coordination, Identity, and Learning},
journal = {Organization Science},
year = {1996},
volume = {7},
number = {5},
pages = {502--518},
doi = {10.1287/orsc.7.5.502},
url = {https://doi.org/10.1287/orsc.7.5.502}
}

@article{grant1996,
author = {Grant, Robert M.},
title = {Toward a Knowledge-Based Theory of the Firm},
journal = {Strategic Management Journal},
year = {1996},
volume = {17},
number = {S2},
pages = {109--122},
doi = {10.1002/smj.4250171110},
url = {https://doi.org/10.1002/smj.4250171110}
}

@book{argote1999,
author = {Argote, Linda},
title = {Organizational Learning: Creating, Retaining and Transferring Knowledge},
publisher = {Kluwer Academic Publishers},
address = {Boston, MA},
year = {1999}
}

@book{nonakaTakeuchi1995,
author = {Nonaka, Ikujiro and Takeuchi, Hirotaka},
title = {The Knowledge-Creating Company: How Japanese Companies Create the Dynamics of Innovation},
publisher = {Oxford University Press},
address = {New York},
year = {1995}
}

@article{nonakaKonno1998,
author = {Nonaka, Ikujiro and Konno, Noboru},
title = {The Concept of ``Ba'': Building a Foundation for Knowledge Creation},
journal = {California Management Review},
year = {1998},
volume = {40},
number = {3},
pages = {40--54},
doi = {10.2307/41165942},
url = {https://doi.org/10.2307/41165942}
}

@article{nonakaVonKrogh2009,
author = {Nonaka, Ikujiro and von Krogh, Georg},
title = {Tacit Knowledge and Knowledge Conversion: Controversy and Advancement in Organizational Knowledge Creation Theory},
journal = {Organization Science},
year = {2009},
volume = {20},
number = {3},
pages = {635--652},
doi = {10.1287/orsc.1080.0412},
url = {https://doi.org/10.1287/orsc.1080.0412}
}

@techreport{chatterji2025people,
  title={How People Use ChatGPT},
  author={Chatterji, Aaron and Cunningham, Thomas and Deming, David J and Hitzig, Zoe and Ong, Christopher and Shan, Carl Yan and Wadman, Kevin},
  year={2025},
  institution={National Bureau of Economic Research}
}

@techreport{idetalamas2026,
author = {Ide, Enrique and Talamàs, Eduard},
title = {The Turing Valley: How AI Capabilities Shape Labor Income},
institution = {IESE Business School},
type = {Working Paper},
month = {January},
year = {2026}
}

@techreport{beraja2026value,
author = {Beraja, Martin and Talamàs, Eduard},
title = {The Value of Organizational Learning Technologies},
institution = {IESE Business School},
type = {Working Paper},
month = {March},
year = {2026},
}

@article{brynjolfsson2025generative,
title={Generative AI at work},
author={Brynjolfsson, Erik and Li, Danielle and Raymond, Lindsey},
journal={The Quarterly Journal of Economics},
volume={140},
number={2},
pages={889--942},
year={2025},
publisher={Oxford University Press}
}

@techreport{johnston2026shift,
author = {Johnston, Drew and Holtz, David and Richmond, Alex Martin and Ong, Christopher and Tambe, Prasanna and Chatterji, Aaron},
title = {The Shift to Agentic {AI}: Evidence from {Codex}},
institution = {OpenAI},
year = {2026},
month = {June},
type = {Working Paper},
url = {https://cdn.openai.com/pdf/5d1e1489-21c0-43e4-9d42-f87efdbf0082/the-shift-to-agentic-ai-evidence-from-codex.pdf}
}

@inproceedings{ouyangEtAl2022,
author = {Ouyang, Long and Wu, Jeff and Jiang, Xu and Almeida, Diogo
and Wainwright, Carroll L. and Mishkin, Pamela and Zhang, Chong
and Agarwal, Sandhini and Slama, Katarina and Ray, Alex
and Schulman, John and Hilton, Jacob and Kelton, Fraser
and Miller, Luke and Simens, Maddie and Askell, Amanda
and Welinder, Peter and Christiano, Paul and Leike, Jan
and Lowe, Ryan},
title = {Training Language Models to Follow Instructions with Human Feedback},
booktitle = {Advances in Neural Information Processing Systems},
volume = {35},
pages = {27730--27744},
year = {2022},
url = {https://cdn.openai.com/papers/Training_language_models_to_follow_instructions_with_human_feedback.pdf}
}

@misc{klowdenTao2026,
author = {Klowden, Tanya and Tao, Terence},
title = {Mathematical Methods and Human Thought in the Age of {AI}},
year = {2026},
note = {Working Paper},
eprint = {2603.26524},
archiveprefix= {arXiv},
primaryclass = {cs.AI},
doi = {10.48550/arXiv.2603.26524},
}

@article{ideTalamas2025,
author = {Ide, Enrique and Talamàs, Eduard},
title = {Artificial Intelligence in the Knowledge Economy},
journal = {Journal of Political Economy},
year = {2025},
volume = {133},
number = {12},
pages = {3762--3800},
doi = {10.1086/737233},
}

@incollection{chatterjiRockTalamas2025,
author = {Chatterji, Aaron and Rock, Daniel and Talamàs, Eduard},
title = {Transformative {AI} and Firms},
booktitle = {The Economics of Transformative {AI}},
editor = {Agrawal, Ajay K. and Brynjolfsson, Erik and Korinek, Anton},
publisher = {University of Chicago Press},
address = {Chicago},
year = {2025},
chapter = {7},
}

@book{polanyi1966,
author = {Polanyi, Michael},
title = {The Tacit Dimension},
publisher = {Doubleday},
address = {Garden City, NY},
year = {1966},
}

@article{nonaka1991,
author = {Nonaka, Ikujiro},
title = {The Knowledge-Creating Company},
journal = {Harvard Business Review},
year = {1991},
volume = {69},
number = {6},
pages = {96--104},
}

@article{nonaka1994,
author = {Nonaka, Ikujiro},
title = {A Dynamic Theory of Organizational Knowledge Creation},
journal = {Organization Science},
year = {1994},
volume = {5},
number = {1},
pages = {14--37},
doi = {10.1287/orsc.5.1.14},
}

@article{brynjolfssonMitchell2017,
author = {Brynjolfsson, Erik and Mitchell, Tom},
title = {What Can Machine Learning Do? Workforce Implications},
journal = {Science},
year = {2017},
volume = {358},
number = {6370},
pages = {1530--1534},
doi = {10.1126/science.aap8062},
}

@article{brynjolfssonMitchellRock2018,
author = {Brynjolfsson, Erik and Mitchell, Tom and Rock, Daniel},
title = {What Can Machines Learn, and What Does It Mean for Occupations and the Economy?},
journal = {AEA Papers and Proceedings},
year = {2018},
volume = {108},
pages = {43--47},
doi = {10.1257/pandp.20181019},
}

@techreport{autor2024,
author = {Autor, David H.},
title = {Applying {AI} to Rebuild Middle Class Jobs},
institution = {National Bureau of Economic Research},
type = {Working Paper},
number = {32140},
year = {2024},
doi = {10.3386/w32140},
}
\end{document}